\documentstyle[11pt,a4]{article}

\newcommand{\vs}[1]{\rule[- #1 mm]{0mm}{#1 mm}}

\newcommand{\be}{\begin{equation}}
\newcommand{\ee}{\end{equation}}

\newcommand{\bea}{\begin{eqnarray}}
\newcommand{\eea}{\end{eqnarray}}

\newcommand{\lp}{\left(}
\newcommand{\rp}{\right)}
\newcommand{\rar}{\rightarrow}

\newcommand{\scr}{\scriptstyle}

\newcommand{\om}{\omega}
\newcommand{\Om}{\Omega}
\newcommand{\al}{\alpha}

\newcommand{\alb}{\bar{\alpha}}
\newcommand{\xb}{\bar{x}}
\newcommand{\pb}{\bar{p}}
\newcommand{\qb}{\bar{q}}
\newcommand{\bs}{\bar{s}}
\newcommand{\eps}{\epsilon}
\newcommand{\PP}{\phi^{\dagger}\phi}
\newcommand{\dV}{\frac{d}{dV}}
\newcommand{\dxi}{\frac{dx_i^2}{dV}(p)}
\newcommand{\dxj}{\frac{dx_j^2}{dV}(p)}
\newcommand{\dxt}{\frac{dx_i^2}{dT}}
\newcommand{\parV}{\frac{\partial}{\partial V}}
\newcommand{\cI}{\oint_{\cal C}\frac{d\om}{4\pi i}}
\newcommand{\Vp}{V^{\prime}}
\newcommand{\Vt}{\tilde{V}}

\newcommand{\pho}{\phi^{(0)}}

\newcommand{\sect}[1]{\setcounter{equation}{0}\section{#1}}

\begin{document}

\begin{titlepage}

\leftline{\Large {CPT-97/P.3455}}
\rightline{\Large {January 1997}}
\leftline{\Large {hep-th/9702005}}

\vs{20}

\begin{center}

{\LARGE {\bf Universal correlators for multi-arc \\[.5cm]
             complex matrix models}}\\[2cm]

{\large G. Akemann} 
\footnote{supported by European Community grant no. ERBFMBICT960997} \\[.5cm]
{\em Centre de Physique Th\'eorique}
\footnote{Unit\'e Propre de Recherche 7061},
{\em CNRS,\\
Case 907 Campus de Luminy, 13288 Marseille, Cedex 9, France}\\
{akemann@cpt.univ-mrs.fr}\\[.5cm]

\end{center} 

\vs{20}

\centerline{ {\bf Abstract}}
The correlation functions of the multi-arc complex matrix model
are shown to be universal for any finite number of arcs. The universality
classes are characterized by the support of the eigenvalue density
and are conjectured to fall into the same classes as the ones 
recently found for the hermitian model. This is explicitly shown
to be true for the case of two arcs, apart from the known result
for one arc.
The basic tool is the iterative solution of the loop equation 
for the complex matrix model with multiple arcs, which provides
all multi-loop correlators up to an arbitrary genus.
Explicit results for genus one are given for any number of arcs.
The two-arc solution is investigated in detail, including the 
double-scaling limit. In addition universal expressions for the string
susceptibility are given for both the complex and hermitian model.

\end{titlepage}

\renewcommand{\thefootnote}{\arabic{footnote}}
\setcounter{footnote}{0}

\sect{Introduction}

\indent

The notion of universality is one of the crucial properties
of random matrix theory. It ensures, that physical quantities of interest 
such as correlation functions or the free energy and its derivatives
- the string susceptibility - do not depend explicitly on the matrix potential
and in this sense are universal. The full dependence on the coupling
constants  can be encoded in few universal parameters as the endpoints
of the support of the spectral density.

Universality can be found in different ways when performing the 
large-$N$ limit. First, in the so-called microscopic limit the 
oscillatory behavior of the eigenvalue density 
is studied and found to be universal \cite{BZ93} (for recent progress
see \cite{KF97} and references therein). In this respect 
particularly the study of the unitary ensembles, the complex and hermitian
one-matrix model, has drawn considerable attention for its application
to 3- and 4-dimensional QCD \cite{VS93}.
There, the merging of the eigenvalue density from two into a single arc 
provides an effective model for the formation of a chiral 
condensate in terms of random matrix theory.

In the second way of performing the large-$N$ limit, the macroscopic limit,
where the oscillations of the eigenvalue density are smoothed, the splitting
of the support into several pieces has been much further understood.
After the seminal works for the one-arc case by Ambj{\o}rn and his 
collaborators for the hermitian and complex matrix model \cite{AJM90,AKM92,
AMB93}, it has been recently shown for the hermitian model,
that for any number of arcs all correlation functions are universal
and can be classified completely by the support of the spectral density
\cite{AA96,AKE96}.

The aim of the present paper is to extend the previous results to the 
complex matrix model with multiple arcs, generalizing the work of \cite{AKM92}.
The same achievements will be made here, obtaining again a whole set
of universal correlators characterized by the support of the density.
Moreover the new universality classes found here are conjectured to coincide
precisely with the ones previously found for the hermitian model \cite{AA96,
AKE96}. This result was already known for one arc \cite{AMB93} and is now 
proven explicitly for two arcs as well. The basic tool therefor will be again
the iterative solution of the loop equation. Furthermore the
string susceptibility of both, the complex and the hermitian model
with two arcs will be calculated explicitly and shown to be universal.

This opens the possibility to investigate the relationship between multi-arc
matrix models and integrable hierarchies using loop equation techniques
\cite{DVV91,MAK91}. The outcome will have to be compared to earlier
results for the two-arc solution from
orthogonal polynomials \cite{HOL91}, which depend heavily
on the reliability of the ansatz for the recursion coefficients. For the
case of two non-symmetric arcs this is already very doubtful \cite{BDJT93}.

Another earlier attempt towards the universality of the two-arc phase 
of the hermitian matrix model has been
made in \cite{MHK95}. However, the completely independent treatment
of the two arcs leads to an expression for the correlator of densities
which fails to fulfill the correct analyticity properties when compared to the 
two-arc eigenvalue density.

\newpage

It should be mentioned that macroscopic universality has been proven
for orthogonal and symplectic ensembles as well, using variational methods
in the saddle-point approximation \cite{Been94}. This result for the correlator
of densities has been extended to planar multi-loop correlators in the 
framework of loop equations \cite{IT96}. However, the appearance of odd 
powers in $1/N$ in the expansion make it very difficult to achieve explicit 
results for higher genera.

Furthermore renormalization group techniques for matrix models
have been recently extended to the multi-arc case as well
\cite{HINS96}. For the hermitian matrix model with a symmetric double-well
potential the authors of \cite{HINS96} find a second attractive fixed
point of the renormalization group transformation, apart from the
Gaussian one. These results confirm the universality of the planar
two-loop correlator for two symmetric arcs by completely different means. 

The present paper is organized as follows.
After briefly introducing the basic definitions in section 2, 
section 3 deals with the planar solution of the loop equations for
the multi-arc complex matrix model. Section 4 contains the iterative
solution of the loop equation, where explicit results are given for genus one
for any number of arcs.
The question of the equivalence with the universality classes of the 
hermitian model is also addressed here. In section 5 the two-arc solution is
presented in more detail including a proof for 
the matching with the corresponding
two-arc universality class of the hermitian model. 
Section 6 contains all the possibilities for performing the double-scaling 
limit for two arcs, 
where the generic case maps to the double-scaled one-arc solution 
\cite{AMB93}. Finally, before 
concluding in the last section the string susceptibility for both
the complex and the hermitian two-arc model is explicitly shown 
to be universal as 
well. Throughout the paper the notation for the solution of the one-arc
complex model \cite{AKM92} will be followed closely.

\sect{Basic Definitions}

\indent

The complex one-matrix model is defined by the partition function
\bea
Z \ [N,\{g_i\}] \ \equiv \ e^{N^2 F} &\equiv& 
\int d\phi^{\dag} d\phi \ \mbox{exp} (-N \ \mbox{Tr} V(\PP)) \ , \nonumber\\
V(\PP) \ &\equiv& \ \sum_{j=1}^\infty \frac{g_j}{j}(\PP)^j \ , \label{Z}
\eea
where the integration is over complex $N \times N$ matrices. The generating 
functional or one-loop correlator is given by
\be
W(p) \ \equiv \ 
\frac{1}{N} \sum_{k=0}^\infty \frac{\langle \mbox{Tr}(\PP)^k \rangle}{p^{2k+1}}
  \ = \ \frac{1}{N} \left\langle \mbox{Tr} \frac{p}{p^2-\PP} \right\rangle \ .
\label{We}
\ee
Introducing the loop insertion operator 
\be
\dV(p) \ \equiv \ - \sum_{j=1}^\infty \frac{j}{p^{2j+1}} \frac{d}{dg_j} 
\ee
the one-loop correlator can be obtained from the free energy,
\be
W(p)\ = \ \dV (p) F\ +\ \frac{1}{p} \ . \label{dF1}
\ee
More generally one gets the $n$-loop correlator by iteratively applying
the loop insertion operator to $W(p)$ (or $F$),
\bea
W(p_1,\ldots,p_n) &\equiv&  N^{n-2} \sum_{k_1,\ldots,k_n=1}^\infty 
\frac{\langle \mbox{Tr} (\PP)^{k_1} \cdots \mbox{Tr}(\PP)^{k_n}\rangle_{conn}}
{p_1^{2k_1+1} \cdots p_n^{2k_n+1}}    \label{Wdef}\\ 
& = & \dV(p_n)\dV(p_{n-1})\cdots\dV(p_2) W(p_1) \ , \ \ n \ge 2 \ , 
\label{dF}
\eea
where $conn$ refers to the connected part.
All the multi-loop correlators and the free energy have the same $1/N$ 
expansion 
\bea
W(p_1,\ldots,p_n) &=& \sum_{g=0}^\infty \frac{1}{N^{2g}} W_g(p_1,\ldots,p_n)
                                                        \label{Wg} \ ,\\ 
F &=&  \sum_{g=0}^\infty \frac{1}{N^{2g}} F_g \ ,
\eea
so relation (\ref{dF}) holds for each genus separately. In particular eq. 
(\ref{dF1}) may be written as 
\be
W_g(p) \ =\ \dV (p)\ F_g \ \ , \ g\geq 1\ \ . \label{Wg1} 
\ee

\sect{The Loop Equation} \label{loopE}

\indent
The loop equation for the multi-arc complex matrix model \cite{AKE95}
looks the same as for one arc \cite{AKM92} apart from the integration contour
$\cal C$ ,
\be 
\cI \frac{\om V^{\prime}(\om)}{p^2-\om^2} W(\om) \ = \ 
   (W(p))^2 + \frac{1}{N^2}\dV(p)W(p) \ \ , \ p\not \in \sigma \ .
   \label{loop}   
\ee
Here $\cal C$ encloses all the square root 
singularities (cuts) of $W(p)$ along the arcs of
support $\sigma$ of the eigenvalue density $\rho(y)$,

\be
\sigma \ = \ \left\{
       \begin{array}{rl}
       \bigcup_{i=1}^{\frac{s}{2}}([-x_{2i-1},-x_{2i}]\cup [x_{2i},x_{2i-1}]) 
                                              & \ \ s  \ \mbox{even} \\
     \Big( \bigcup_{i=1}^{\frac{s-1}{2}} ([-x_{2i-1},-x_{2i}] \cup 
        [x_{2i},x_{2i-1}])\Big) \cup [-x_s,x_s] & \ \ s \ \mbox{odd .}
       \end{array}  \right. \label{sig}
\ee
where $x_1> \ldots >x_s>0$ . The integration contour ${\cal C}$
can thus be depicted in the following way

\newpage

\begin{figure}[h]
\unitlength1cm
\begin{picture}(12.2,5.5)
\linethickness{1.0mm}
\multiput(2.1,2.3)(6,0){2}{\line(1,0){2.4}}
\thinlines
\put(0,2.3){\line(1,0){1.9}}
\put(4.7,2.3){\line(1,0){1}}
\multiput(6,2.3)(0.3,0){3}{\circle*{0.05}}
\put(6.9,2.3){\line(1,0){1}}
\put(10.7,2.3){\vector(1,0){2}}
\multiput(2,2.3)(2.6,0){2}{\circle*{0.2}}
\multiput(8,2.3)(2.6,0){2}{\circle*{0.2}}
\put(5.8,3.5){\circle*{0.2}}
\put(6.1,3.7){$p$}
\put(1.7,2){$-x_1$}
\put(4.3,2){$-x_2$}
\put(8,2){$x_2$}
\put(10.6,2){$x_1$}
\multiput(2.,2.3)(6,0){2}{\oval(1.0,2.3)[l]}
\multiput(4.6,2.3)(6,0){2}{\oval(1.0,2.3)[r]}
\multiput(2.,3.45)(6,0){2}{\line(1,0){1.2}}
\multiput(3.4,1.15)(6,0){2}{\line(1,0){1.2}}
\multiput(4.6,3.45)(6,0){2}{\vector(-1,0){1.4}}
\multiput(2.,1.15)(6,0){2}{\vector(1,0){1.4}}
\put(1.2,3.3){${\cal C}_{s}$} 
\put(7.2,3.3){${\cal C}_{1}$} 
\end{picture} 
\caption{
        $ {\cal C} =  \cup_{i=1}^s {\cal C}_{i} $ }
                     \label{fig2}
\end{figure}

As it has been explained already in \cite{DJM92} the eigenvalue density can be
introduced either for real values $y$ or for real positive values $y^2$,
$y\in$ R, where $y^2$ is the eigenvalue of $\PP$. In the first convention, 
which is chosen throughout the calculations, $\rho(y)$ is a smooth,
symmetric function on R, whereas in
the latter it becomes singular when having support at the origin
\cite{DJM92}. In eq. (\ref{sig}) this is just the case when $s$ is {\it odd}.
The corresponding density on $\mbox{R}_+$ is obtained by transforming
$\rho(y)\rightarrow \rho(y)/2y\equiv\rho(y^2)$.

Now there is a simple way to see why the case of an {\it even} $s$ is actually 
equivalent to the hermitian model with $s/2$ cuts on $\mbox{R}_+$\footnote{
On the level of the eigenvalue model in \cite{DJM92} this is intuitively
clear.}. When identifying 
\bea
\frac{1}{p} W (p) & \rar & W^H(\pb)\Big|_{\pb=p^2} \ , \nonumber\\
(\pm x_i)^2 &\rar& \xb_i >0 \ \ , \ i=1,\ldots,s \ , \label{Wequiv}
\eea
one exactly arrives at the loop equation of the hermitian 
matrix model ({\it H}) with $s/2$ cuts
\cite{AKE96}, plus an additional term $\sim 1/\pb$. This term must vanish
since $W^H(\pb )$ has to be regular at $\pb =0$ for $s$ {\it even},
as the origin is not included in the support $\sigma$ then\footnote{
This fact had been missed in \cite {AKE95}, giving a simple 
explanation why the results presented there for $s=2$ are equivalent to 
\cite{AMB93}.}. In \cite{DJM92} the same observation had been made on the 
level of orthogonal polynomials for the case of $s=2$, identifying the
respective string equations of the complex and hermitian model. 
Consequently in all
the following only the case of $s$ {\it odd} will be further considered.
It will be called $(s+1)/2$-arc solution, counting the number of arcs on
$\mbox{R}_+$. This way of counting is chosen to name the same universality 
class in the hermitian and complex model by the same number of arcs.

Solving the loop equation (\ref{loop}) 
in the planar limit the $1/N^2$-term
can be omitted. The result is given by \cite {AKE95} 
\be
W_0(p) \ = \ \frac{1}{2}\cI\frac{wV^{\prime}(\om)}{p^2-\om^2}
             \sqrt{\prod_{i=1}^s\frac{p^2-x_i^2}{\om^2-x_i^2}} \ ,
\label{W0}
\ee
which can be rewritten as 
\be
W_0(p) \ = \ \frac{1}{4}\lp \Vp(p)-M(p)
             \sqrt{\prod\nolimits_{i=1}^s(p^2-x_i^2)}\rp \ . \label{WM}
\ee
Here $M(p)$ is an analytic function given by
\be
M(p) \ = \ \oint_{\cal C_{\infty}} \frac{d\om}{4\pi i}
\frac{2\om\Vp(\om)}{\om^2-p^2}\frac{1}{\sqrt{\prod_{i=1}^s(\om^2-x_i^2)}} \ . 
\label{Mfin}
\ee
The signs of the square root in the complex plain in eq. (\ref{WM})
are chosen to be $(\pm p)^s$ at $\pm \infty$.
From eq. (\ref{WM}) the eigenvalue density $\rho(y)$ 
can be immediately read off:
\bea
\rho(y) &\equiv& \lim_{\eps\to 0}\frac{1}{2\pi i}
        \lp W_0(y-i\eps)\ -\ W_0(y+i\eps)\rp \ \ ,\ y\in\sigma\ , \ \nonumber\\
 &=& \frac{1}{4\pi}\left| M(y)\right|
          \sqrt{-\prod\nolimits_{i=1}^s(y^2-x_i^2)} \ .
\label{rho}
\eea
The parameters $x_i$, $i=1,\ldots,s$, given as functions of the coupling
 constants $g_i$, $i\in \mbox{N}_+$, 
are determined by the boundary conditions on
$W_0(p)$. From the asymptotic behavior $W(p)\sim\frac{1}{p}$ eq. 
(\ref{We}) it follows
\be
\delta_{l,s} \ = \ \frac{1}{2} 
 \cI \om^l \Vp(\om)\frac{1}{\sqrt{\prod_{i=1}^s(\om^2-x_i^2)}} 
 \ \ ,\ l=1,3,\ldots,s \ \ .
 \label{bc1}
\ee
In analogy to the hermitian case \cite{AKE96} the remaining $\frac{s-1}{2}$ 
equations for the $x_i$
are obtained by a criterion of stability \cite{DAV90}\footnote{Because 
of symmetry the stability criterion has to be imposed only between
half of all arcs.} 
\be
0 \ =\   \int_{x_{2k+1}}^{x_{2k}}dy M(y) 
                     \sqrt{\prod\nolimits_{j=1}^s(y^2-x_j^2)} \ \ , 
        \ k=1,\ldots,\frac{s-1}{2} \ \ .\label{bc2}
\ee
For $s\geq 3$ these equations lead to the explicit 
appearance of elliptic integrals
in the higher genus correlators as well as in the planar $n$-loop
correlators for $n\geq 2$.

\sect{The iterative solution}\label{iter}

\indent

After having solved the loop equation (\ref{loop}) 
in leading order, an iterative
scheme can be built up for calculating higher order contributions. It is based
on the planar solution (\ref{W0}) 
along the same lines as in \cite{AKM92}, which is treating the one-cut 
situation. Inserting the genus expansion of the one-loop correlator
eq. (\ref{Wg}) into the loop eq. (\ref{loop}) 
and comparing order by order in genus,
the higher order contributions are determined by
\be
(\hat{{\cal K}}-2W_0(p))
W_g(p) = \sum_{g\prime=1}^{g-1}W_{g\prime}(p)
      W_{g-g\prime}(p)+\dV (p)W_{g-1}(p) \ , \ g\ge 1 \ .
\label{loopg}
\ee
Here $\hat{{\cal K}}$ is a linear integral operator, given by
\be
\hat{{\cal K}} f(p) \equiv \cI \frac{\om \Vp(\om)}{p^2-\om^2}f(\om) \ .
\ee
Once the result for $W_0(p)$ is known one can determine $W_g(p)$ for $g\geq 1$
iteratively from contributions of lower genus on the r.h.s. of eq. 
(\ref{loopg}),
provided that the operator $(\hat{{\cal K}}-2W_0(p))$ can be uniquely inverted.
However, it can be shown from the boundary conditions (\ref{bc1}), that this 
operator possesses the following zero modes
\bea
\mbox{Ker}(\hat{{\cal K}}-2W_0(p)) &=& \mbox{Span}
 \{ \pho(p),p^2\pho(p),\ldots,p^{s-1}\pho(p)\} \ ,\nonumber\\
\pho (p)&\equiv& \frac{1}{\sqrt{\prod_{i=1}^s(p^2-x_i^2)}}  \ . 
\label{zero}
\eea
Because of the asymptotic of $W_g(p)\sim {\cal O}(\frac{1}{p^3})$ for
$g\geq 1$, which can be seen from the definition (\ref{We}),
the last zero mode $\sim \frac{1}{p}$
cannot contribute to $W_g(p)$. In the case of $s=1$
this argument uniquely fixes the inversion of eq. (\ref{loopg}) \cite{AKM92}.
Here, the condition (\ref{Wg1}) will be used in addition to achieve a
unique solution $W_g(p)$ for any $s$. Therefore a set of basis functions
is introduced in the following way:
\bea
(\hat{{\cal K}}-2W_0(p))\ \chi_i^{(n)}(p) \ &=&\ \frac{1}{(p^2-x_i^2)^n} \ , 
                        \ i=1,\ldots,s\ \ , \ n\in \mbox{N}_+ \ ,\nonumber\\
(\hat{{\cal K}}-2W_0(p))\ \psi^{(n)}(p) \ &=&\  \frac{1}{p^{2n}} \ 
                       \ \ \ \ \ \ \ ,\ n\in \mbox{N}_+\ . \label{basisdef}
\eea
In addition a change of variables from the coupling constants $g_i$ to the
following moments is performed:
\bea
I_i^{(k)} \ &\equiv&\   \cI \frac{\om\Vp(\om)}{(\om^2-x_i^2)^k} \pho (\om)
        \  , \ \ i=1,\ldots,s\ , \ k\in \mbox{N}_+ \ ,\nonumber\\
 M^{(n)} &\equiv& \cI \frac{\Vp(\om)}{\om^{2n+1}} \pho (\om)
        \ \ \ \ \ \ , \ \ n\in \mbox{N} \ . \label{mom}
\eea
Apart from the endpoints of the cuts $x_i$ these moments will encode the 
universality of the higher genus correlators. Namely given that the r.h.s.
of eq. (\ref{loopg}) is a fractional rational function of $p$ having poles at
the $x_i$ only, $W_g(p)$ will have the following structure:
\be
W_g(p) \ =\ \sum_{n=1}^{3g-1}\sum_{i=1}^s A_{i,g}^{(n)}\chi_i^{(n)}(p)\ +\ 
            \sum_{n=1}^{g}D_{g}^{(n)}\psi^{(n)}(p) \ .
\label{Wgsol}
\ee
The coefficients of the basis functions $A_{i,g}^{(n)}$ and $D_{g}^{(n)}$ 
are complicated functions of the endpoints $x_i$ 
and of the moments eq. (\ref{mom}). As the order of the highest
pole in $W_g(p)$ is not changed compared to the one-cut case \cite{AKM92},
$W_g(p)$ will depend on at most $s(3g-1)$ moments $I^{(k)}_i $and $g$
moments $M^{(n)}$.
This follows from the same 
arguments as in \cite{AKM92}.
The unique set of basis functions can be proven to be 
(see appendix \ref{Abasis}):
\bea
\chi_i^{(n)}(p) &\equiv& \frac{1}{I_i^{(1)}}\lp
   \phi_i^{(n)}(p)\Big|_{\frac{d}{dV}-part}
   - \sum_{k=1}^{n-1}\chi_i^{(k)}(p)I_i^{(n-k+1)} \rp \ ,\nonumber\\
\psi^{(n)}(p) &\equiv& \frac{1}{M^{(0)}}\lp
   \Om^{(n)}(p)\Big|_{\frac{d}{dV}-part}
    - \sum_{k=1}^{n-1}\psi^{(k)}(p)M^{(n-k)} \rp \ ,\nonumber\\
&&\phi_i^{(n)}(p) \equiv \frac{1}{(p^2-x_i^2)^n}\pho(p) \ ,\ \  
i=1,\ldots,s \ \ ,\ n\in\mbox{N}_+ \ ,\nonumber\\ 
&&\Om^{(n)}(p) \equiv \frac{1}{p^{2n}}\pho(p) \ ,\ \ \ \ \ \ \ \ 
n\in\mbox{N}_+ \ .\label{basis}
\eea
The restriction to the part which can be written as a derivative of the 
loop insertion operator $\dV(p)$ ensures that the whole $p$-dependence
of the basis functions and thus of $W_g(p)$ in eq. (\ref{Wgsol}) can be
absorbed into such derivatives in order to fulfill eq. (\ref{Wg1}).
At the same time this requirement uniquely fixes the ambiguities coming from 
the zero modes in eq. (\ref{zero}).

The explicit construction of the basis functions can be found in appendix A,
the first of which reading 
\bea
\chi_i^{(1)}(p) &=& \frac{1}{x_i^2}\dxi \ , \ \ \ \ \ \ \ i=1,\ldots,2s
    \ ,\nonumber\\
\chi_i^{(2)}(p) &=& \frac{-2}{3x_i^2I_i^{(1)}}
           \frac{dI_i^{(1)}}{dV}(p) 
 -\frac{2}{3x_i^4}\dxi 
 -\frac{1}{3x_i^2}\sum^{s}_{j=1 \atop j\neq i}\frac{1}{x^2_j-x^2_i}
\Big( \dxj -\dxi\Big)  \nonumber\\
\psi^{(1)}(p) &=& -\frac{2}{M^{(0)}}\frac{dM^{(0)}}{dV}(p)-
\sum^{s}_{i=1}\frac{1}{x^2_i}\dxi \ .
\label{basis1}
\eea
The one-loop correlator of genus one $W_1(p)$ can now be calculated from
eq. (\ref{loopg}) for $g=1$, which reads
\be
(\hat{\cal K}-2W_0(p))W_1(p) \ = \ \dV(p)W_0(p) \ . \label{loop1} 
\ee
Therefore the planar two-loop correlator $W_0(p,p)$
on the r.h.s. has to be determined, 
which can be achieved from applying $\dV(p)$ to eq. (\ref{W0}) in the following
form
\bea
\dV (p)& =&\parV (p) + \sum_{i=1}^{s}\Big( 
     \dxi \frac{\partial}{\partial x_i^2} 
  +\sum_{k=1}^{\infty}\frac{dI_i^{(k)}}{dV}(p) 
          \frac{\partial}{\partial I_i^{(k)}} \Big)
+ \sum_{k=0}^{\infty}\frac{dM^{(k)}}{dV}(p) 
          \frac{\partial}{\partial M^{(k)}} \nonumber\\
&&\parV (p) \equiv - \sum_{j=1}^\infty \frac{j}{p^{j+1}} 
                      \frac{\partial}{\partial g_j} \ . \label{dVsum}
\eea
Using the results from the appendix A eq. (\ref{dxsol}) one obtains
\bea
W_0(p,p) &=&\frac{1}{16}\sum_{i=1}^{s}\frac{x_i^2}{(p^2-x_i^2)^2}     
                  -\frac{1}{16}\sum_{i,j=1 \atop i\neq j}^{s}
                 \frac{x_i^2}{x_i^2-x_j^2}\lp\frac{1}{p^2-x_i^2}
                 -\frac{1}{p^2-x_j^2}\rp\nonumber\\
           &&+\ \frac{1}{16p^2}\ 
              -\ \frac{s}{16}\sum_{i=1}^{s}\frac{1}{p^2-x_i^2} 
             \ +\ \frac{1}{4}\sum_{i=1}^{s}
          \sum_{l=0}^{\frac{s-3}{2}}\frac{\al_{i,l}\ x_i^{2l}}{p^2-x_i^2} \ ,
\label{W0pp}
\eea
where the $\al_{i,l}$ are given by the solution of the linear set of equations 
(\ref{alsyst}). They depend only on the endpoints of the cuts $x_i$ and on 
elliptic functions of them. The result for $W_1(p)$ now follows from inverting
eq. (\ref{loop1}) by inserting the basis functions eq. (\ref{basis1})
into $W_0(p,p)$:
\bea
W_1(p)&=&  \frac{1}{16}\sum_{i=1}^{s}x_i^2 \chi_i^{(2)}(p)     
                 \ -\ \frac{1}{16}\sum_{i,j=1 \atop i\neq j}^{s}
               \frac{x_i^2}{x_i^2-x_j^2}\lp\chi_i^{(1)}(p)-\chi_j^{(1)}(p) \rp
              \nonumber\\
           &+&\ \frac{1}{16}\psi^{(1)}(p) 
           -\frac{s}{16}\sum_{i=1}^{s}\chi_i^{(1)}(p)  
            +\frac{1}{4}\sum_{i=1}^{s}
          \sum_{l=0}^{\frac{s-3}{2}}\al_{i,l}\ x_i^{2l}\chi_i^{(1)}(p) . 
\label{W1}
\eea
The calculation of higher genera $W_g(p)$ as well as the explicit integration
to get the free energy $F_g$ becomes very technical as one has to differentiate
or to integrate the set of elliptic functions given in eq. (\ref{dx}).
Still, following the arguments from \cite{AKM92} it becomes clear, that
the general structure for $W_g(p)$ will be precisely the one given in eq. 
(\ref{Wgsol}) for an arbitrary number of cuts.

In the case $s=3$ the elliptic integrals occurring are given in terms of 
the well known complete
elliptic integrals of first, second and third kind (see eq. (\ref{ell}))
making more explicit results accessible. In the generic double-scaling limit 
the solution will be mapped to the double-scaled one-cut 
solution \cite{AMB93}. This relation 
is probably true for all multi-cut solutions. 
In \cite{AMB93} higher genus results have been
obtained explicitly up to and including $g=4$.

Apart from the general result for genus one for any $s$ also the planar
two-loop correlator at different arguments can be calculated explicitly,
using again the results from appendix A. 
Performing a similar calculation as for $W_0(p,p)$ it follows
\bea
W_0(q,p) &=& \frac{1}{4}\frac{\pho (q)}{\pho (p)} \Bigg(
              \frac{2q^2}{(q^2-p^2)^2}
          +\frac{1}{q^2-p^2}\Big(s-1+\sum_{i=1}^{s}\frac{x_i^2}{p^2-x_i^2}\Big)
              \nonumber\\
           && \ \ \ \ \ \ \ \ \ \ +\sum_{i=1}^{s}\sum_{l=0}^{\frac{s-3}{2}}
          \frac{\al_{i,l}\ q^{2l}}{p^{2}-x_i^{2}} \Bigg)
             \ -\ \frac{1}{2}\frac{pq}{(q^{2}-p^{2})^2} \ \ .\label{W0pq}
\eea
Thus $W_0(p,q)$ and because of eqs. (\ref{dF}) and (\ref{dVsum}) 
also all the higher planar
multi-loop correlators are universal in the sense that they
only depend on the endpoints of the cuts $x_i$. The same statement can be
made for the density correlators, as they can be obtained from the
multi-loop correlators and vice versa (see e.g. \cite {AA96}). Consequently 
eq. (\ref{W0pq}) gives a classification of all universality classes
of the multi-arc complex matrix model in terms of the support of the 
eigenvalue density.

A similar classification has been found in previous works
for the hermitian model \cite{AA96,AKE96}, 
so a natural question to ask is whether
one can identify the classes here and there, which have the same number
of intervals as support. In order to do so one has to map again the correlation
functions from the hermitian model \cite{AKE96} to the corresponding quantities
in the complex model as in eq. (\ref{Wequiv}),
\be
\frac{1}{p^2}W_0(p,p) \ \to \ W_0^H(\pb,\pb)\Big|_{\pb=p^2} \ .
\label{Wppequiv}
\ee
In addition the smallest endpoint $\xb_{2\bs}$
of the cuts of the hermitian model $(H)$ has to be set
to zero, as now the complex model with 
{\it odd s} is treated\footnote{In matching the correlators of the two models
the density for nonnegative eigenvalues of the complex 
model is considered.}.
While for the one-cut case it is known already, that the universality classes
for the planar two-loop correlators match \cite{AJM90}, in the general case
$W_0(p,p)$ from eq. (\ref{W0pp}) has to be compared with \cite{AKE96}
\bea
\pb\ W^H_0(\pb,\pb)
             &=&\frac{1}{16}\sum_{i=1}^{2\bs-1}\frac{\xb_i}{(\pb-\xb_i)^2}     
                 -\frac{1}{16}\sum_{i,j=1 \atop i\neq j}^{2\bs-1}
                 \frac{\xb_i}{\xb_i-\xb_j}\lp\frac{1}{\pb-\xb_i}
                 -\frac{1}{\pb-\xb_j}\rp  \nonumber\\
           &+&\frac{1}{16\pb}
             - \frac{2\bs-1}{16}\sum_{i=1}^{2\bs-1}\frac{1}{\pb-\xb_i} 
             +\frac{1}{4}\sum_{i=1}^{2\bs-1}
   \sum_{l=0}^{\frac{2\bs-4}{2}}\frac{\alb_{i,l}\xb_i^{l}}{\pb-\xb_i} 
             + \frac{\al_{2\bs,0}}{4\pb}
\label{Wh0pp}
\eea
for $\pb=p^{2}$, $\xb_i=x_i^{2}$, $i=1,\ldots,2\bs-1$ and $\xb_{2\bs}\to 0$. 
When identifying $s=2\bar{s}-1$, which is the corresponding number of 
endpoints, the two correlators eq. (\ref{W0pp}) and eq. (\ref{Wh0pp})
are identical apart from the last extra term in eq. (\ref{Wh0pp})
which is non vanishing 
in the limit $x_{2\bar{s}}\to 0$. However, one has to bare in mind
that the $\alb_{i,l}$ are solutions to a set
of $2\bs (\bs-1)$ linear equations (see eq. (5.16) in \cite{AKE96}),
which are different from the $s(s-1)/2$ eqs. (\ref{alsyst}) 
determining the $\al_{i,l}$. A proof for the equivalence
of the two sets of universality classes would thus need some nontrivial 
identities for the elliptic integrals occurring and is therefore left
as a conjecture. Still, for the case of $s=3$ this proof can performed 
explicitly as it will be shown in the next section.
Summing up the results obtained so far
the complex model with both $s$ even {\it and} odd
contains all universality classes found for the unitary ensembles, 
due to the fact that the hermitian model is contained in the complex one 
for the case of $s$  even.

\sect{The two-cut solution $s=3$}\label{2cut}

\indent

As it has been mentioned already
for $s=3$ further results become accessible as the elliptic integrals 
originating from the boundary conditions can be related to the complete 
elliptic integrals of the first, second and third kind. 
Explicit expressions for the zero mode coefficients $\al_{i}$ of the 
quantities $\dxi$,
which are responsible for the appearance of the elliptic integrals, are given 
in appendix B eq. (\ref{alf}).

In the previous section the set of universality classes for all multi-arc
solutions for the complex model has been given and an equivalence to the 
ones recently found in the hermitian model \cite {AA96,AKE96} was conjectured.
In the special case considered
this equivalence will be shown to hold.
For $s=3$ the general results obtained for $W_0(p,p)$ eq. 
(\ref{W0pp}) and $W_0(p,q)$ eq. (\ref{W0pq}) take the following form, using
the results for the $\al_i$ from appendix B,

\newpage

\bea
W_0(p,p)&=& \frac{1}{16}\sum_{i=1}^3\frac{x_i^2}{(p^2-x_i^2)^2}
         +\frac{1}{16p^2} 
- \frac{1}{16}\sum^3_{i<j}\frac{x_i^2+x_j^2}{(p^2-x_i^2)(p^2-x_j^2)}\nonumber\\
  &&+\frac{1}{16}\frac{1}{p^2-x_1^2}
   +\frac{3(x_2^2+x_3^2)-2p^2}{16(p^2-x_2^2)(p^2-x_3^2)}
      +\frac{1}{4}\frac{x_2^2(x_1^2-x_3^2)}{\prod_{j=1}^3(p^2-x_j^2)}
        \frac{E(k)}{K(k)}   \nonumber\\ \\
W_0(p,q)&=&\frac{1}{4(p^2-q^2)^2}\Bigg(
 p^2 \sqrt{\frac{(p^2-x_1^2)(q^2-x_2^2)(q^2-x_3^2)}{(p^2-x_2^2)(p^2-x_3^2)
  (q^2-x_1^2)}}
\nonumber\\
&&\ \ \ \ \ \ \ \ \ \ \ \ \ \ \ \  
   +\ q^2\sqrt{\frac{(p^2-x_2^2)(p^2-x_3^2)(q^2-x_1^2)}{(p^2-x_1^2)(q^2-x_2^2)
(q^2-x_3^2)}}
 \Bigg)\nonumber\\
      &+&\frac{1}{4}\ 
\frac{x_2^2(x_1^2-x_3^2)}{\sqrt{\prod_{j=1}^3(p^2-x_j^2)(q^2-x_j^2)}}
        \frac{E(k)}{K(k)} 
     \ -\ \frac{1}{2}\frac{pq}{(p^2-q^2)^2} \ .
\eea
The consistency of the two results can be checked explicitly. Namely
$W_0(p,q)$ is {\it symmetric}, it has the same {\it analyticity} 
properties as $W_0(p)$, it {\it asymptotically} reaches 
$\sim{\cal O}(\frac{1}{p^3})$, it is {\it regular} at 
equal arguments and coincides with $W_0(p,p)$ and it finally has the correct
{\it discontinuity} $-\frac{pq}{(p^2-q^2)^2}$ along the cuts.
The conjectured equivalence of the universality class
of the hermitian model with two cuts at $[\xb_4,\xb_3]\cup [\xb_2,\xb_1]$ 
is now evident when comparing with the results given in \cite{AKE96} (eqs. 
(6.6) and (6.5)). 
Setting $\xb_4 =0$ and identifying $\xb_i=x_i^2$ it holds
\bea
\frac{1}{p^2}W_0(p,p) &\equiv& W_0^H(\pb,\pb)\Big|_{\pb=p^2\atop \xb_i=x_i^2,\
 \xb_4=0} \ , \nonumber\\
\frac{1}{pq}W_0(p,q) &\equiv& W_0^H(\pb,\qb)\Big|_{\pb=p^2,\ \qb=q^2 \atop
 \xb_i=x_i^2,\ \xb_4=0} \ .
\eea
So despite the fact that in the complex model the eigenvalue density 
$\rho(y^2)$ becomes singular at the origin whereas the density of the hermitian
model stays smooth, both models belong to the same universality class
for one and two cuts. Nevertheless, 
for higher genera the two models will give in general
different results, as it can be seen for example 
from comparing the free energy obtained below with the one in \cite{AKE96}.
It is only in the double-scaling limit,
that all the models mentioned will become completely 
equivalent (see sect. \ref{DSL}).

The correlator of densities given in \cite{MHK95}, which can be related to
the planar two-loop correlator
$W_0(p,q)$ (see e.g. \cite{AA96}), does not match with the above
two-arc universality class for the following reasons. First of all the authors
of \cite{MHK95} do not impose any boundary condition as in eq. (\ref{bc2})
and are therefore left with a parameter dependent expression. Furthermore
their independent treatment of the two arcs, adding up two single arc
correlators with different endpoints, does not fulfill the correct
analyticity properties compared to the two-arc eigenvalue density in 
\cite{AKE96}.

Turning back to the results for genus $g=1$ from the previous section also
the free energy $F_1$ can now be determined. This is due to the fact that the
derivatives of the complete elliptic integrals with respect to their moduli
are known and close among themselves (see e.g. \cite{Byrd}). From 
eq. (\ref{W1}) the one-loop correlator of genus one for $s=3$ 
takes the following form,
\bea
W_1(p)&=&-\frac{1}{24}\sum_{i=1}^3\Big(
    \frac{1}{I_i^{(1)}}\frac{dI_i^{(1)}}{dV}(p)+\frac{1}{x_i^2}\dxi 
    +\sum_{j> i}\frac{1}{x_j^2-x_i^2}\big( \dxj -\dxi\big) \Big)
       \nonumber\\
    &&-\ \frac{1}{16}\sum_{i\neq j}\frac{1}{x_i^2-x_j^2}
           \Big(\dxi - \frac{x_i^2}{x_j^2}\dxj\Big) 
      \ - \frac{1}{8}\frac{1}{M^{(0)}}\frac{dM^{(0)}}{dV}(p) \nonumber\\
    &&-\ \frac{1}{4}\sum_{i=1}^3\frac{1}{x_i^2}\dxi 
      \ +\ \frac{1}{4}\sum_{i=1}^3\frac{\al_i}{x_i^2}\dxi \ .
\label{W1two}
\eea
Using the results for the $\al_i$ eq. (\ref{alf}) and standard formula 
for the complete elliptic integrals \cite{Byrd} $F_1$ can be integrated 
according to eq. (\ref{Wg1})
\bea
F_1 &=& -\frac{1}{24}\sum_{i=1}^3\ln |I_i^{(1)}|-\frac{1}{8}\ln |M^{(0)}|
        -\frac{1}{2}\ln |K(k)| -\frac{1}{6}\sum_{i=1}^3\ln |x_i^2|\nonumber\\
    && -\frac{1}{6}\sum_{i < j}\ln |x_i^2-x_j^2|
      +\frac{1}{4}\lp \ln |x_1^2-x_3^2| + \ln |x_2^2|\rp \ .
\label {F1}
\eea
Apart from the factor of the $M^{(0)}$-term all the other terms coincide
with the free energy of genus one of the hermitian model \cite{AKE96} at
$\xb_4=0$. The same observation can be made when comparing the free
energies of the one-cut solutions. There, at higher genera the difference 
between the hermitian and complex model becomes more evident.
The iterative procedure of solving the loop equation is purely algebraic
and in the case of $s=3$ even accessible to computer  algebra, grace to the
knowledge of the elliptic integrals. In the next section higher genera
will be obtained by other means, namely by comparing to the results
for the one-cut solution in the double-scaling limit \cite{AMB93}.


\sect{The double-scaling limit for $s=3$}\label{DSL}

\indent

In general two different situations may occur in the double-scaling
limit (d.s.l.). In the generic case, when the limit is performed at any
of the endpoints $\pm x_i$, the contributions coming from the
multi-cut structure will drop out and the correlators will precisely
match with the continuum version of the one-cut case \cite{AMB93}.
A different situation is given, when the d.s.l. is performed at
simultaneously shrinking or/and merging arcs. The correlators will differ
from the one-arc result, as it has been found similarly in the d.s.l.
of the hermitian two-cut model \cite{AKE96}. 

\subsection{The scaling limit at $x_j$}

Choosing any endpoint, say $x_j$, the $m$-th multi-critical point 
is defined by $m-1$ extra zeros of the eigenvalue density eq. (\ref{rho})
accumulating at $+x_j$ and $-x_j$. By tuning the coupling constants in order
to reach this point the argument of the correlators $p$ and the $\pm x_j$
will scale like 
\bea
p^2 &=& (x_j^c)^2 \ +\ a\pi \ \ , \nonumber\\
x_j^2 &=& (x_j^c)^2 \ -\ a\Lambda^{\frac{1}{m}} \ \ ,
\label{gdsl}
\eea
with $a\to 0$, whereas the $x^2_{i\neq j}$ do not scale. Since
$I_i^{(k)}=\frac{1}{2}\frac{1}{(k-1)!}
\frac{d^{k-1}M(p)}{dp^{2(k-1)}}\Big|_{p=x_i}$
the $j$-th moments will scale like
\be
I_j^{(k)}\ \sim \ a^{m-k} \ \ ,
\ee
whereas the other moments $I_{i\neq j}^{(k)}$ and $M^{(n)}$
do not scale either. In the limit eq. (\ref{gdsl}) the elliptic integrals
$K(k)$ and $E(k)$ and consequently the zero mode contributions $\al_i$
will stay subdominant in eqs. (\ref{dxsol}), (\ref{dI}) and (\ref{dM}),
\bea
\dxj &=& \frac{x_j^2}{I_j^{(1)}}\phi_j^{(1)}(p) \ \ \ \ \mbox{(d.s.l.)}
\ , \label{dxdsl} \\
\frac{dI_j^{(k)}}{dV}(p)&=& 
  (k+\frac{1}{2}) \lp I_j^{(k+1)} \dxj - x_j^{2}\phi_j^{(k+1)}(p) \rp
  \ \ \ \ \mbox{(d.s.l.)}\ .
\label{dxdI}
\eea
Only the basis functions $\chi_j^{(k)}(p)\sim a^{-m-n+\frac{1}{2}}$
will give dominant contributions when inverting the loop equation.
Moreover the loop insertion operator itself will simplify due to 
eq. (\ref{dxdI}) as well as $W_0(p,p)$ in eq. (\ref{W0pp}) , 
the starting point of the iteration,
\bea
W_0(p,p)&=& \frac{1}{16}\frac{x_j^2}{(p^2-x_j^2)^2} \ \ \ \ \mbox{(d.s.l.)}\ ,
\nonumber\\
\dV (p) &=&\dxj \frac{\partial}{\partial x_j^2} 
           \ +\ \sum_{k=1}^{\infty}\frac{dI_j^{(k)}}{dV}(p) 
          \frac{\partial}{\partial I_j^{(k)}}  \ \ \ \ \mbox{(d.s.l.)} \ .
\label{WdV}
\eea
The eqs. (\ref{dxdsl}) -- (\ref{WdV}) precisely map to the corresponding 
quantities of the double-scaled one-cut solution, which are given in 
the appendix A of \cite{AMB93}. Consequently both solutions are equivalent 
in any order of genus, as they start the iteration from the same quantities.
In fact more was proven in \cite{AMB93}, namely that in the d.s.l. the 
complex and the hermitian matrix model become completely equivalent.
From the above results this equivalence now extends to the complex model for
$s=3$. The same matching has been shown for the two-cut hermitian model 
\cite{AKE96}. This equivalence has to be understood in the symbolical sense 
that the basic objects as
moments and endpoints of support appear exactly in the same way.
However, when expanding these objects in terms of coupling 
constants $g_i$, the expansion explicitly depends on the number of
cuts.

\subsection{The scaling limit at merging and shrinking arcs}

In the case where the d.s.l. is considered at an endpoint of an arc,
which is simultaneously shrinking to zero, merging with another arc,
or both together, the following situations are possible:

\begin{itemize}
\item[i)] $x_2\to x_3\ \ $ merging 
\item[ii)] $x_3\to \ 0\ \ $  shrinking
\item[iii)] $x_2\to x_1\ \ $ shrinking 
\item[iv)] $x_2,x_3\to \ 0$ merging + shrinking 
\item[v)] $x_1,x_2\to x_3$ merging + shrinking 
\end{itemize}

\begin{figure}[h]
\unitlength1cm
\begin{picture}(12.2,2.5)
\linethickness{1.0mm}
\multiput(2.1,1.3)(6,0){2}{\line(1,0){2.4}}
\put(5.9,1.3){\line(1,0){0.8}}
\thinlines
\put(0,1.3){\line(1,0){1.9}}
\put(4.7,1.3){\line(1,0){1}}
\put(6.9,1.3){\line(1,0){1}}
\put(10.7,1.3){\vector(1,0){2}}
\multiput(2,1.3)(2.6,0){2}{\circle*{0.2}}
\multiput(8,1.3)(2.6,0){2}{\circle*{0.2}}
\multiput(5.8,1.3)(1.0,0){2}{\circle*{0.2}}
\put(1.6,0.8){$-x_1$}
\put(4.2,0.8){$-x_2$}
\put(8.0,0.8){$x_2$}
\put(10.5,0.8){$x_1$}
\put(5.4,0.8){$-x_3$}
\put(6.7,0.8){$x_3$}
\end{picture} 
\caption{The support $\sigma$ of the eigenvalue density $\rho(y)$ for $s=3$
                     \label{fig3}}
\end{figure}
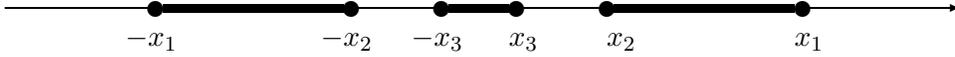

In all five cases the resulting continuum limit will differ from the previous
one called generic, and thus from the d.s.l. of the one-cut solution.
Only in the last case v) the contributions containing elliptic integrals in 
eqs. (\ref{dxsol}), (\ref{dI}) and (\ref{dM}) will stay dominant. In all five
cases the iterative solution of the loop equation with not simplify much
any longer.
The case i) will be treated in more detail as an example,
whereas in the remaining cases only the scaling behavior of the zero mode
coefficients $\al_i$ of the quantities $\dxi$ will be given.
\begin{itemize}
\item[i)] The transition of merging cuts can be parameterized by
\bea
x_2^2 &=& x_c^2 \ +\ a\nu \ , \nonumber\\
x_3^2 &=& x_c^2 \ -\ a\mu \ , \ \ \nu,\mu>0 \ ,\nonumber\\
p^2 &=& x_c^2 \ +\ a\pi \ .\label{dsl1}
\eea
In this limit the modulus $k$ of the elliptic integrals vanishes,
\be
k^2\ =\ a(\nu+\mu)\frac{x_1^2}{x_c^2(x_1^2-x_c^2)} \ +\ {\cal O}(a^2) \ ,
\ee
and consequently $K(k)$ and $E(k)$ can be expanded in terms of $k^2$,
\be
\frac{E(k)}{K(k)}\ =\ 1-\frac{1}{2}k^2  \ +\ {\cal O}(a^2) \ .
\ee
The coefficients $\al_i$ will stay finite in the limit (\ref{dsl1}),
\bea
\al_1 &=& \frac{x_1^2}{x_1^2-x_c^2}  \ +\ {\cal O}(a) \ , \nonumber\\
\al_2 &=& \frac{1}{2}- \frac{x_c^2}{2(x_1^2-x_c^2)}  \ +\ {\cal O}(a) \ ,
\nonumber\\
\al_3 &=& \al_2  \ +\ {\cal O}(a)  \ ,
\eea
leaving the zero mode contributions $\al_i \pho(p)\sim a^{-1}$ subdominant
compared to $\phi_{j=2,3}^{(1)}\sim a^{-2}$ in eq. (\ref{dxtwo}). 
Due to the scaling
of both sets of moments $I_{j=2,3}^{(k)}$ their derivative now reads
\bea
\frac{dI_2^{(k)}}{dV}(p)&=& (k+\frac{1}{2})\lp I_2^{(k+1)}\frac{dx_2^2}{dV}(p)
                             -x_2^2\phi_2^{(k+1)}(p)\rp \nonumber\\
           &&+\ \frac{1}{2}\sum_{l=1}^k\frac{1}{(x_3^2-x_2^2)^{k-l+1}}
             \lp x_3^2\phi_2^{(l)}(p)-I_2^{(l)}\frac{dx_3^2}{dV}(p)\rp \ 
\mbox{(d.s.l.)}\ , \nonumber\\
\label{dI1}
\eea
and the same for the indices 2 and 3 interchanged. The starting point for 
the iteration $W_0(p,p)$ becomes
\be
W_0(p,p)\ =\ \frac{1}{16}\sum_{i=2,3}\frac{x_i^2}{(p^2-x_i^2)}
             -\frac{1}{16}\frac{(x_2^2+x_3^2)}{(x_2^2-x_3^2)}
             \lp \frac{1}{p^2-x_2^2}
              - \frac{1}{p^2-x_3^2}\rp \ \mbox{(d.s.l.)} \label{W0dsl1}
\ee
from which $W_1(p)$ can be easily obtained. Clearly the result differs 
from the generic d.s.l. in the previous subsection.

\item[ii)]The d.s.l. at the inner arc
shrinking to zero can be parameterized as
\bea
x_3^2 &=& a \ , \nonumber\\
p^2   &=& a\pi \ .
\eea
As the modulus $k^2$ reaches unity in this limit the elliptic integrals 
can be expanded in terms of the complementary modulus $k^{\prime 2}$
\bea
k^{\prime 2} &\equiv& 1-k^2 \ =\ a\ \frac{x_1^2-x_2^2}{x_1^2x_2^2} \nonumber\\
\Rightarrow \ \frac{E(k)}{K(k)}&\sim& {\cal O}(\frac{1}{\ln a})\ .
\eea
The $\al_i$ will stay finite or vanish,
\bea
\al_1 &=& \al_2 \ =\ 1\ +\ {\cal O}(\frac{1}{\ln a}) \ ,\nonumber\\
\al_3 &\sim& {\cal O}(\frac{1}{\ln a}) \ .
\eea
Still, the continuum limit will be changed compared to the generic d.s.l.
as now the moments $M^{(k)}$ will also scale.
\item[iii)] The shrinking of the outer arcs is parameterized by
\bea
x_1^2 &=& x_c^2 \ +\ a\nu \ , \nonumber\\
x_2^2 &=& x_c^2 \ -\ a\mu \ , \ \ \mu,\nu>0\ ,\nonumber\\
p^2 &=& x_c^2 \ +\ a\pi \ . \label{dsl3}
\eea
Expanding again in the complementary modulus $k^{\prime 2}$ leads to
\be
\frac{E(k)}{K(k)}\ \sim \ {\cal O}(\frac{1}{\ln a}) 
\ee
for the elliptic integrals and hence to the following behavior for the 
zero mode coefficients
\bea
\al_1,\al_2 &\sim& {\cal O}(\frac{1}{a\ln a}) \ , \nonumber\\
\al_3 &=& -\frac{x_3^2}{x_c^2-x_3^2} \ + \ {\cal O}(\frac{1}{\ln a}) \ .
\eea
Although they become singular in the limit (\ref{dsl3})
they still remain subdominant, leading again to eq. (\ref{dxdsl}) for $j=1,2$.
The starting point for the iteration becomes like eqs. (\ref{dI1}) and 
(\ref{W0dsl1}), with the index 3 replaced by 1.
\item[iv)]The merging of the arcs while the inner arc is shrinking to zero is 
para\-meterized by
\bea
x_2^2 &=& a\nu \ , \nonumber\\
x_3^2 &=& a\mu \ ,\ \nu>\mu>0 \ ,\nonumber\\
p^2 &=& a\pi \ .
\eea
In this limit the modulus reaches
\be
k^2 \ =\ 1-\frac{\mu}{\nu} \ +\ {\cal O}(a) \ ,
\ee
which is not necessarily small and may take any value in (0,1). 
The $\al_i$ become
\bea
\al_1 &=& 1\ + \ {\cal O}(a) \ ,\nonumber\\
\al_2 &=& \frac{\nu}{\nu-\mu}\lp  1-\frac{E(k)}{K(k)}\rp\ + \ {\cal O}(a) 
\ ,\nonumber\\
\al_3 &=& \frac{1}{\nu-\mu}\lp -\mu+\nu\frac{E(k)}{K(k)}\rp\ +\ {\cal O}(a)\ .
\eea
Consequently they do not contribute in the d.s.l..
\item[v)]In the case of merging arcs while the outer arcs are shrinking to zero
the $\al_i$ will be sufficiently enhanced. The d.s.l. reads
\bea
x_1^2 &=& x_c^2\ +\ a\nu \ ,\nonumber\\
x_2^2 &=& x_c^2\ +\ a\mu \ ,\ \ \nu>\mu>0 \ ,\nonumber\\
x_3^2 &=& x_c^2 \ ,\nonumber\\
p^2   &=& x_c^2\ +\ a\pi \ . 
\eea
In this limit the modulus becomes 
\be
k^2 \ =\ \frac{\mu}{\nu} \ +\ {\cal O}(a) \ ,
\ee
which can reach any value in (0,1). The zero mode coefficients now read
\bea
\al_1 &=& a^{-1}\ \frac{x_c^2}{(\nu-\mu)}\frac{E(k)}{K(k)} \ +\ {\cal O}(a^0) 
 \ ,\nonumber\\
\al_2 &=& a^{-1}\ \frac{x_c^2}{\mu}
         \lp 1-\frac{\nu}{(\nu-\mu)}\frac{E(k)}{K(k)}\rp \ +\ {\cal O}(a^0) \ ,
     \nonumber\\
\al_3 &=& a^{-1}\ \frac{x_c^2}{\mu}\lp -1+\frac{E(k)}{K(k)}\rp  \ +\ 
      {\cal O}(a^0) \ .
\eea
In this case all $\al_i$ become sufficiently singular to contribute in the
$\dxi$ in the d.s.l.. The iterative procedure will therefore not simplify
and in addition the elliptic integrals remain explicitly present in the 
continuum limit.
This interesting new continuum behavior has also been found in the 
hermitian model \cite{AKE96}.
\end{itemize}

\sect{The string susceptibility}

\indent

In this section a closed and universal expression will be given 
for the string susceptibility of both the complex and hermitian 
matrix model with two cuts. It reflects once more the fact that
the two-point correlators of both models are universal as well. 
As the calculation for the complex 
model is much simpler it is presented in some detail. 
For the hermitian model only the final result will be given,
which can be obtained along the same lines when starting from \cite{AKE96}.
Both susceptibilities are shown to belong to the same class of 
universality. The knowledge of the string susceptibility may either serve
to study the critical behavior of the model, as for example 
in \cite{EK95} for the $O(n)$-model or in \cite{AKM94} for the Penner
model.
On the other hand when one tries to relate matrix models to integrable
hierarchies using loop equation techniques \cite{DVV91,MAK91}
the susceptibility enters directly the corresponding Gelfand-Dikii ansatz.

In order to introduce an overall coupling constant in front of the potential
the following replacement is performed,
\be
V(p)\ \to\ \frac{1}{T}\Vt (p) \ .
\label{VVT}
\ee
The parameter $T$ can be thought of as a temperature or the cosmological
constant. The string susceptibility $U(T)$ is then obtained from the 
planar free energy $F_0$,
\bea
U(T) &=& \frac{d^2}{dT^2} \lp T^2F_0 \rp  \nonumber\\
     &\sim& (T_c-T)^{-\gamma_{str}}  \ .\label{Udef}
\eea
Here it is supposed, that $U(T)$ shows some critical behavior at
$T=T_c$. In order to exploit the knowledge of the quantities calculated
so far as $W_0(p)$, the two following identities will be used, which can be
shown easily from the definitions,
\bea
T^2\frac{dF_0}{dT} &=& 2\cI W_0(\om) \Vt (\om) \ , \nonumber\\
T\frac{dU}{dT} &=& \frac{d^2}{dT^2}\lp T^3\frac{dF_0}{dT} \rp 
\ =\ \frac{d^2}{dT^2}\lp 2\cI T W_0(\om) \Vt (\om)\rp \ . \label{dUdF}
\eea
The final result obtained will thus be $\frac{dU}{dT}$.
Taking $W_0(p)$ from eq. (\ref{W0}) together with the replacement 
(\ref{VVT}) the first differentiation of the integrand yields 
\be
\frac{d}{dT}(TW_0(\om)) \ =\ \frac{1}{4}\sum_{i=1}^s T I_i^{(1)}\dxt
    \frac{1}{(\om^2-x_i^2)\pho (\om)}  \ . \label{dtW}
\ee
In order to proceed the quantities $\dxt$ have to be determined,
which can be done exactly along the same lines as the calculation
of the $\dxi$. Here $s=3$ has to be specified to obtain explicit results.
Differentiating the boundary equations (\ref{bc1}) and (\ref{bc2}) with 
respect to $T$ leads to the following set of equations
\bea
0&=& \sum_{i=1}^3 T I_i^{(1)}\dxt \ ,\nonumber\\
4&=& \sum_{i=1}^3 x_i^2T I_i^{(1)}\dxt \ ,\nonumber\\
0&=& \sum_{i=1}^3 T I_i^{(1)}\dxt K_i\ .\label{dxtdet}
\eea
The solution is obtained with the help of appendix B eq. (\ref{ell}),
where the $K_i$ are given,
\be
T I_i^{(1)}\dxt \ =\ \frac{4}{\prod_{j\neq i}(x_i^2-x_j^2)}
                 \lp x_i^2 - x_3^2\frac{\Pi(\al,k)}{K(k)} \rp 
       \ ,\ \ i=1,2,3\ .\label{dxt}
\ee
Inserting this into eq. (\ref{dtW}) for $s=3$ one arrives after some 
calculation at 
\be
\frac{d}{dT}(TW_0(\om)) \ =\ \pho (\om)
           \lp \om^2 -x_3^2\frac{\Pi(\al,k)}{K(k)} \rp \ ,
\ee
and consequently
\be
T\frac{dU}{dT}\ =\ \frac{d}{dT} 
    \lp 2\cI\pho (\om)\Big( \om^2-x_3^2\frac{\Pi(\al,k)}{K(k)}\Big) \Vt (\om)
                       \ \rp \ . \label{dt2}
\ee
As the differentiation of the complete elliptic integrals with respect
to their moduli is known (e.g. in \cite{Byrd}), it is useful to 
rewrite $\frac{d}{dT}$ in the following way
\be
\frac{d}{dT}\ =\       \frac{\partial}{\partial T}  
 + \sum_{i=1}^{s}\dxt \frac{\partial}{\partial x_i^2} 
+   \frac{dk^2}{dT} \frac{\partial}{\partial k^2} 
 +   \frac{d\al^2}{dT} \frac{\partial}{\partial \al^2} \ .\label{dTsum}
\ee
Using this form on the r.h.s. of eq. (\ref{dt2}) one arrives again after some
tedious calculation at the following result for the integrand
\bea
&&\frac{d}{dT}\lp \pho (\om) (\om^2 -x_3^2\frac{\Pi(\al,k)}{K(k)})\rp \ =\ 
\nonumber\\
&&=\frac{1}{2}\sum_{i=1}^3\dxt
       \lp 1-\frac{x_3^2}{x_i^2}\frac{\Pi(\al,k)}{K(k)}\rp 
                       \lp x_i^2\phi_i^{(1)}(\om)+\al_{i}\pho(\om)\rp \ .
\label{dt3}
\eea
In order to get a universal expression for $\frac{dU}{dT}$ in eq. 
(\ref{dt2}) one would like to 
perform a partial integration to arrive at an integral containing $\Vp (\om)$.
Next the boundary conditions eqs. (\ref{bc1}) and (\ref{bc2}) would allow
to eliminate the explicit dependence on the potential. Indeed the wanted
transformation can be done due to the following observation. 
The last parenthesis in eq. (\ref{dt3}) containing the $\om$-dependence
is precisely given by $I_i^{(1)}\frac{dx_i^2}{dV}(\om)$, which can be
written as a total derivative with respect to $\om$. This becomes evident when
rewriting the linear set of equations (\ref{dbcB}), which determine these 
quantities\footnote{The same can be achieved for the general set of equations
(\ref{dxA}) and (\ref{dx}).},
\bea
0&=& \sum_{i=1}^3 I_i^{(1)}\dxi \ +\ \frac{\partial}{\partial p}(p\pho(p))
\ , \nonumber\\
0&=& \sum_{i=1}^3 x_i^2 I_i^{(1)}\dxi \ +\ 
       \frac{\partial}{\partial p}(p^3\pho(p)) \ , \nonumber\\
0&=& \sum_{i=1}^3 I_i^{(1)}\dxi K_i\ -\ 
       \frac{\partial}{\partial p}(p\pho(p)A(p)) \ , \nonumber\\
&&A(p) \ =\ \int_{x_3}^{x_2}dy\frac{1}{p^2-y^2}\frac{1}{\pho(y)}
\label{dxw}
\eea
The last equation can be shown to hold using eq. (\ref{dx}). In the same way 
the boundary condition (\ref{bc2}) is shown to be equivalent to 
\be
0\ =\ \cI \Vp (\om)\ 2\om\pho (\om) A(\om) \ . \label{bc2n}
\ee
From eq. (\ref{dxw}) one obtains
\be
I_1^{(1)}\frac{dx_1^2}{dV}(p)=\frac{1}{\Sigma}\frac{\partial}{\partial p}
  \Big( (K_2-K_3)p^3\pho(p)+ (x_2^2-x_3^2)p\pho(p) + 
        (x_2^2-x_3^2)p\pho(p)A(p) \Big) \label{dxdw}
\ee
where
\be
\Sigma \ =\ (x_2^2-x_3^2)K_1 +(x_3^2-x_1^2)K_2 +(x_1^2-x_2^2)K_3 \ .
\ee
The $\dxi$ for $i=2,3$ are obtained by a cyclic permutation
of the indices. When inserting eq. (\ref{dxdw}) into 
eq. (\ref{dt3}), integrating by parts in eq. (\ref{dt2}) 
and using the boundary conditions (\ref{bc1}) and (\ref{bc2n}) 
only the terms $\sim p^3\pho(p)$ will survive in eq. (\ref{dxdw}). The result 
for the string susceptibility finally reads
\be
\frac{dU}{dT}\ =\ \sum_{i=1}^3\dxt \frac{2}{\prod_{j\neq i}(x_i^2-x_j^2)}
                 \lp x_i^2\ -\ x_3^2\frac{\Pi(\al,k)}{K(k)}\rp^2 \ .
\label{dUC}
\ee
It is a remarkable universal quantity, as it shows no explicit potential 
dependence. Compared to the one-arc susceptibility 
\bea
\frac{dU_{s=1}}{dT} &=& \frac{2}{x^2}\frac{dx^2}{dT}\ , \nonumber\\
\frac{dx^2}{dT} &=& \frac{4}{TI^{(1)}} \ ,
\eea
which can be obtained as a simple exercise, eq. (\ref{dUC}) contains much more
structure and thus offers more possibilities for showing critical behavior.

The string susceptibility for the hermitian matrix model 
with two cuts can be calculated exactly along the same lines presented in this
section, taking \cite{AKE96} as a starting point. It is given by
\be
\frac{dU^H}{dT}\ =\ \sum_{i=1}^4\frac{d\xb_i}{dT} 
                \frac{2}{\prod_{j\neq i}(\xb_i-\xb_j)}\lp \xb_i-\xb_4 
         \ +\ (\xb_4-\xb_3)\frac{\Pi(\bar{\al},\bar{k})}{K(\bar{k})}\rp^2 \ ,
\label{dUH}
\ee
where 
\be
T \bar{I}_i^{(1)}\frac{d\xb_i}{dT} \ =\ \frac{4}{\prod_{j\neq i}(\xb_i-\xb_j)}
 \lp \xb_i-\xb_4\ +\ (\xb_4-\xb_3)\frac{\Pi(\bar{\al},\bar{k})}{K(\bar{k})}\rp
, \ i=1,\ldots,4 \ .
\ee
To compare with the result for the complex model eq. (\ref{dUC}) $\xb_4$ is set
to zero, as being done already in section \ref{iter}. It can be seen now 
explicitly, that both string susceptibilities belong to the same universality
class. This is not so astonishing, since the planar two-loop correlators 
of the complex and hermitian two-cut model have been identified already
in section \ref{2cut}. For completeness the one-arc susceptibility is also 
given for the hermitian model,
\bea
\frac{dU_{s=1}^H}{dT}&=& \sum_{i=1,2}\frac{d\xb_i}{dT}
                         \frac{2}{\prod_{j\neq i}(\xb_i-\xb_j)}\ ,\nonumber\\
T\bar{I}_i^{(1)}\frac{d\xb_i}{dT} &=& \frac{4}{\prod_{j\neq i}(\xb_i-\xb_j)} 
\ ,\ \ i=1,2\ .
\eea


\sect{Conclusions}

\indent

In this paper the program of classifying and calculating correlation functions
of the unitary ensembles, the hermitian and complex one-matrix model, is
completed.
For any number of arcs of the support of the eigenvalue density the 
correlation functions are found to be universal for both models, thus providing
a complete classification of all possible correlators to occur.
The higher order corrections in $1/N^2$ to the correlators are determined
by an iterative solution of the loop equation, where for genus one explicit
results are obtained for any number of arcs. In general, in 
the double-scaling limit
the one- and two-arc solution of the complex and hermitian model become 
equivalent all together. However, when the scaling limit is performed 
at merging or/and shrinking arcs, a new continuum behavior has been
found here in the complex model as well as in the hermitian model \cite{AKE96}.

An interesting question is to find applications for the multi-arc 
solutions, which were 
calculated here in the macroscopic large-$N$ limit. For example in solid states
physics such situations with a Hamiltonian having a band structure for
its eigenvalues typically occur. 

It seems to be straightforward 
to analyze the critical behavior of the two-arc solutions now directly
from the given string susceptibilities. This could be done in a similar way as 
the investigation of all possible transitions 
in the double-scaling limit in section \ref{DSL}. 
Another point left for future work is the 
relation between the multi-arc matrix models and integrable 
hierarchies. 
The use of loop equation techniques \cite{DVV91,MAK91} requires the
knowledge of the string susceptibility as it enters directly the corresponding
Gelfand-Dikii ansatz. In \cite{HOL91} the mKdV and a subset of the
nonlinear Schr\"odinger (NLS) hierarchy has been proposed for the symmetric
and non-symmetric two-arc solution of the hermitian matrix model. 
However, in the latter case the ansatz made there for the orthogonal 
polynomials is very doubtful \cite{BDJT93}, so
it remains to be checked whether the same results can be obtained from the
loop equations.

\indent

\begin{flushleft}
\underline{Acknowledgments}: 
This work was supported 
by European Community grant no. ERBFMBICT960997. I would like to thank  
J. Ambj{\o}rn, C. Kristjansen and Yu. Makeenko for helpful discussions.
In addition I wish to thank Yu. Makeenko for sharing with me his unpublished
manuscript \cite{MAK91}.
\end{flushleft}


\begin{appendix}

\sect{Derivation of the basis} \label{Abasis}

\indent

It can be easily shown by induction, that the functions given in eq. 
(\ref{basis}) without the restriction to the $\dV$-part form a possible 
set of basis functions, satisfying eq. (\ref{basisdef}) 
up to the addition of allowed zero modes of eq. (\ref{zero}):
\bea
\tilde{\chi}_i^{(n)}(p) &\equiv& \frac{1}{I_i^{(1)}}\lp \phi_i^{(n)}(p)-
                   \sum_{k=1}^{n-1}\tilde{\chi}_i^{(k)}(p)I_i^{(n-k+1)} \rp 
              \ , \ i=1,\ldots,2s \ \ , \ \ n\in \mbox{N}_+\ , \nonumber\\
\tilde{\psi}^{(n)}(p) &\equiv& \frac{1}{M^{(0)}}\lp \Om^{(n)}(p)-
                   \sum_{k=1}^{n-1}\tilde{\psi}^{(k)}(p)M^{(n-k)} \rp
              \ \ \ , \ \ n\in \mbox{N}_+\ . 
\label{basisA}
\eea
In order to absorb the $p$-dependence  of these functions into derivatives
of the loop insertion operator, the $\dV(p)$-part of $\phi_i^{(n)}(p)$
and $\Om^{(n)}(p)$ has to be calculated. The basis functions are then
fixed uniquely by subtracting the remaining part, which will turn out to
consist only of zero modes. Therefore the $\dV$-derivative of the basic 
variables $I_i^{(k)}$, $M^{(k)}$ and $x_i^{2}$ has to be determined.
This can be achieved  by rewriting the loop insertion operator as in eq.
(\ref{dVsum})
and applying it to the definitions (\ref{mom}). Using the identity
\be
\parV (p) \Vp(\om) \ =\ \frac{-2\om p}{(p^{2}-\om^{2})^2} \ \ ,
\ee
the results read
\bea
\frac{dI_i^{(k)}}{dV}(p)&=& 
  (k+\frac{1}{2}) \lp I_i^{(k+1)} \dxi-x_i^{2}\phi_i^{(k+1)}(p) \rp
   \ - \ (k+\frac{s-1}{2})\phi_i^{(k)}(p)  \nonumber\\
  &+&\ \frac{1}{2} \sum_{j=1 \atop j\not= i}^{s}
  \sum_{l=1}^k \frac{1}{(x^2_j-x_i^2)^{k-l+1}} \Big( x_j^{2}\phi_i^{(l)}(p)-
          I_i^{(l)} \frac{dx_j^{2}}{dV}(p) \Big)                  \nonumber\\
  &+&\ \frac{1}{2} \sum_{j=1 \atop j\not= i}^{s}  \frac{1}{(x_j^{2}-x_i^{2})^k}
           \Big( I_j^{(1)}\frac{dx_j^{2}}{dV}(p)-x_j^{2}\phi_j^{(1)}(p) \Big),
   \ i=1,\ldots,2s \label{dI} \\
\frac{dM^{(k)}}{dV}(p)&=& 
   - \frac{2k+1}{2}\Om^{(k+1)}(p) 
  \ -\ \frac{1}{2} \sum_{i=1}^{s}
  \frac{1}{x_i^{2k+2}} \Big( x_i^{2}\phi_i^{(1)}(p)-
          I_i^{(1)} \frac{dx_i^{2}}{dV}(p) \Big)                  \nonumber\\
  &&-\frac{1}{2} \sum_{i=1}^{s} \lp \dxi\sum_{l=0}^k
  \frac{M^{(l)}}{x_i^{2(k-l+1)}}
  \ -\ \sum_{l=1}^k  \frac{1}{x_i^{2(k-l+1)}}\Om^{(l)}(p) \rp  .   \label{dM}
\eea
The $\dxi $ can be obtained from solving a linear set of equations 
derived from the boundary conditions. Applying $\dV (p)$ in the form of eq.
(\ref{dVsum}) to the first set of boundary equations (\ref{bc1})
leads to 
\be
0=\sum_{i=1}^{s}\lp x_i^{k-1} I_i^{(1)}\dxi- 
            x_i^2p^{k-1}\phi_i^{(1)}(p)\rp
     + (k-s)\ p^{k-1}\pho(p) , \  k=1,3,\ldots,s \label{dxA}
\ee
For the second set of boundary 
equations (\ref{bc2}) one gets\footnote{The argument $p$ has to
be excluded to be between the cuts during the 
calculation and analytically continued in the end 
(compare to \cite{AKE96}, appendix A).}
\bea
0&=&\frac{1}{2}\frac{\partial}{\partial p}  
      \int\limits^{x_{2j}}_{x_{2j+1}}dy 
    \frac{2p}{p^2-y^2}\frac{\pho (p)}{\pho (y)} 
        \ -\ \sum_{i=1}^s I_i^{(1)}\dxi K_{i,j} \nonumber\\
&=& \sum_{i=1}^{s}\lp x_i^2 \phi_i^{(1)}(p)\ -\ I_i^{(1)}\dxi \rp K_{i,j}\ ,
    \ \ \ \ \ \ \ \ \ \ j=1,\ldots,\frac{s-1}{2}\ ,  \nonumber\\
&&K_{i,j} \ \equiv \int\limits_{x_{2j+1}}^{x_{2j}}dy
    \frac{\sqrt{\prod_{k=1}^{s}(y^2-x_k^2)}}{(y^2-x_i^2)}\ ,
 \label{dx}
\eea
Making the ansatz 
\be
I_i^{(1)}\dxi \ = \ x_i^2\phi_i^{(1)}(p)\ +\ \sum_{l=0}^{\frac{s-3}{2}}
\al_{i,l}\ p^{2l}\pho (p) 
\ \ , \ \ \ \ \ i=1,\ldots,2s  \ \ ,
\label{dxsol}
\ee
where the sum is over the allowed zero modes in eq. (\ref{zero}),
one arrives at the following set of equations,
\bea
0&=& \sum_{i=1}^{s}\Big(\sum_{l=0}^{\frac{s-3}{2}}\al_{i,l}\ x_i^{2k}\ p^{2l}
     - \sum_{l=0}^{k-1}x_i^{2(k-l)}\ p^{2l}\Big) +(2k+1-s)\ p^{2k},
     \  k=0,\ldots,{\scr \frac{s-1}{2}} \nonumber\\
0&=& \sum_{i=1}^{s}\sum_{l=0}^{\frac{s-3}{2}}\al_{i,l}\ K_{i,j}\ p^{2l}  \ ,
     \ \ j=1,\ldots,{\scr \frac{s-1}{2}} \ . \label{alsyst}
\eea
Comparing powers of $p$ leads to precisely $s(s-1)/2$ linear equations
determining the $\al_{i,l}$, $i=1,\ldots,s$, $l=0,\ldots,\frac{s-3}{2}$, 
completely as functions of $x_i$ and $K_{i,j}$, which are elliptic functions
of the $x_i$. The $\al_{i,l}$ are thus uniquely determined by the support of 
the eigenvalue density $\sigma$. In appendix B they are calculated explicitly
for the example $s=3$.

After having determined the $\dxi$ the $\dV$-part of $\phi^{(k)}_i(p)$
and $\Om^{(k)}(p)$ for the basis can now be obtained as follows. Resolving 
eqs. (\ref{dI}) and (\ref{dM}) for $\phi^{(k+1)}_i(p)$
and $\Om^{(k+1)}(p)$, it becomes evident together with eq. (\ref{dxsol}),
that the part which is not expressed in terms of $\dxi$, 
$\frac{dI_i^{(k)}}{dV}(p)$ or $\frac{dM^{(k)}}{dV}(p)$ is just a linear 
combination of allowed zero modes. These terms can now be
subtracted in eq. (\ref{basisA}) in order to achieve the unique basis
eq. (\ref{basis}). The firsts terms needed for the results in eq. 
(\ref{basis1}) read
\bea
\phi^{(1)}_i(p)\Big|_{\dV -part}&=& \frac{I_i^{(1)}}{x_i^2}\dxi 
\ ,\ \ \ \ i=1,\ldots,s \ \ ,\nonumber\\
\phi^{(2)}_i(p)\Big|_{\dV -part}&=& \frac{I_i^{(2)}}{x_i^2}\dxi 
      -\frac{2}{3x_i^2}\frac{dI_i^{(1)}}{dV}(p)
      - \frac{2I_i^{(1)}}{3x_i^4}\dxi \nonumber\\
      &&-\ \frac{I_i^{(1)}}{3x_i^2}\sum_{j=1 \atop j\not= i}^{s} 
         \frac{1}{x_j^2-x_i^2}\lp \dxj - \dxi\rp
\ ,\nonumber\\
\Om^{(1)}(p)\Big|_{\dV -part}&=& -2 \frac{dM^{(0)}}{dV}(p)
          +\sum_{i=1}^s\frac{1}{x_i^2}\dxi M^{(0)}\ .
\eea

\sect{The $\dxi$ for $s=3$}\label{dxB}

\indent

In this appendix the equations determining the zero mode coefficients
$\al_{i,l}$ of the quantities $\dxi$ 
are explicitly solved for $s=3$. To start with the set
of equations (\ref{dxA}) and (\ref{dx}) for $s=3$ read
\bea
0&=& \sum_{i=1}^{3}\lp I_i^{(1)}\dxi - x_i^2\phi_i^{(1)}(p)\rp -2\pho(p) \ , 
\nonumber\\
0&=& \sum_{i=1}^{3}x_i^2\lp I_i^{(1)}\dxi - x_i^2\phi_i^{(1)}(p)\rp 
-\pho(p)\sum_{i=1}^3x_i^2 \ , \nonumber\\
0&=& \sum_{i=1}^{3}\lp I_i^{(1)}\dxi - x_i^2 \phi_i^{(1)}(p) \rp K_{i}\ ,
\nonumber\\
&&K_{i} \ \equiv \int\limits_{x_3}^{x_2}dy
    \frac{\sqrt{\prod_{k=1}^{3}(y^2-x_k^2)}}{y^2-x_i^2}\ .
\label{dbcB}
\eea
After substituting $y^2\to y$ in the elliptic integrals $K_i$ they can be found
in \cite{Byrd}, where they are expressed by the complete elliptic integrals of
first, second and third kind:
\bea
K_1&=&X\Big( x_2^2x_3^2K(k) + x_2^2(x_3^2-x_1^2)E(k)
              + x_3^2(x_1^2-x_2^2-x_3^2)\Pi(\al,k)\Big)  \nonumber\\
K_2&=&X\Big( (2x_1^2-x_2^2)x_3^2K(k) + x_2^2(x_3^2-x_1^2)E(k)
              + x_3^2(-x_1^2+x_2^2-x_3^2)\Pi(\al,k)\Big)  \nonumber\\
K_3&=&X\Big( (2x_1^2-x_3^2)x_2^2K(k)+x_2^2(x_3^2-x_1^2)E(k)
              + x_3^2(-x_1^2-x_2^2+x_3^2)\Pi(\al,k)\Big) \nonumber\\
&&\nonumber\\
X &=& \frac{1}{2}\frac{1}{\sqrt{x_2^2(x_1^2-x_3^2)}} \ , \ \ 
\al^2\ =\ \frac{x_2^2-x_3^2}{x_2^2}\ , \ \ 
k^2\ =\ \frac{x_1^2(x_2^2-x_3^2)}{x_2^2(x_1^2-x_3^2)} \ .\label{ell}
\eea
The decomposition into $K(k)$, $E(k)$ and $\Pi(\al,k)$ is not unique due to 
possible transformations of the moduli $k$ and $\al$. 
The ansatz (\ref{dxsol}) for $s=3$ reads
\be
I_i^{(1)}\dxi \ =\ x_i^2\phi_i^{(1)}(p) \ +\ \al_i\pho (p)\ , \ \ i=1,2,3 \ ,
\label{dxtwo}
\ee
where $\al_i\equiv\al_{i,0}$. Substituting it into eq. (\ref{dbcB}) 
one obtains the following linear set of equations for the $\al_i$
\bea
0&=&\sum_{i=1}^3 \al_i \ -\ 2 \ ,\nonumber\\
0&=&\sum_{i=1}^3 (\al_i-1)x_i^2 \ ,\nonumber\\
0&=&\sum_{i=1}^3 \al_i K_i \ \ \ .
\label{alfsyst}
\eea
With the help of eq. (\ref{ell}) the solution is obtained, which determines
the $\dxi$ completely
\bea
\al_1 &=& 1\ +\ \frac{x_2^2}{(x_1^2-x_2^2)}\frac{E(k)}{K(k)} \ ,\nonumber\\
\al_1 &=& \frac{x_2^2}{x_2^2-x_3^2}
   \lp 1\ +\ \frac{(x_3^2-x_1^2)}{(x_1^2-x_2^2)}\frac{E(k)}{K(k)}\rp 
\ ,\nonumber\\
\al_1 &=& \frac{x_3^2}{x_3^2-x_2^2}
\lp 1\ -\ \frac{x_2^2}{x_3^2}\frac{E(k)}{K(k)} \rp\ .
\label{alf}
\eea

\end{appendix}

\newpage


\newcommand{\NP}[3]{{\it Nucl. Phys. }{\bf B#1} (#2) #3}
\newcommand{\NPA}[3]{{\it Nucl. Phys. }{\bf A#1} (#2) #3}
\newcommand{\JPA}[3]{{\it J. Phys. }{\bf A#1} (#2) #3}
\newcommand{\PL}[3]{{\it Phys. Lett. }{\bf B#1} (#2) #3}
\newcommand{\PRL}[3]{{\it Phys. Rev. Lett.}{\bf #1} (#2) #3}
\newcommand{\PRD}[3]{{\it Phys. Rev. }{\bf D#1} (#2) #3}
\newcommand{\PRB}[3]{{\it Phys. Rev. }{\bf B#1} (#2) #3}
\newcommand{\IMP}[3]{{\it Int. J. Mod. Phys }{\bf #1} (#2) #3}
\newcommand{\MPL}[3]{{\it Mod. Phys. Lett. }{\bf #1} (#2) #3}

\end{document}